\documentclass[prl,letterpaper,twocolumn,showpacs,superscriptaddress,amsfonts,amssymb]{revtex4}

\usepackage{amsmath}

\newcommand{\bb}[1]{{\mathbb #1}}

\renewcommand{\epsilon}{\varepsilon}
\renewcommand{\phi}{\varphi}

\begin{document}

%
%

\title{Clausius inequality and optimality of quasi static \\
  transformations for nonequilibrium stationary states}

\author{L. Bertini}
\affiliation{Dipartimento di Matematica, Universit\`a di Roma
La Sapienza, Piazza A. Moro 2, 00185 Roma, Italy}

\author{D. Gabrielli}
\affiliation{Dipartimento di Matematica, Universit\`a dell'Aquila,
67100 Coppito, L'Aquila, Italy}

\author{G. Jona-Lasinio}
\affiliation{Dipartimento di Fisica and INFN,
Universit\`a di Roma La Sapienza, Piazza A. Moro 2, 00185 Roma,
Italy}

\author{C. Landim}
\affiliation{IMPA, Estrada Dona Castorina 110, J. Botanico, 22460 Rio
de Janeiro, Brazil, \\ and CNRS UMR 6085, Universit\'e de Rouen,
F76801 Saint-\'Etienne-du-Rouvray, France}


%
%

\begin{abstract}
  Nonequilibrium stationary states of thermodynamic systems dissipate
  a positive amount of energy per unit of time. If we consider
  transformations of such states that are realized by letting the
  driving depend on time, the amount of energy dissipated in an
  unbounded time window becomes then infinite.  Following the general
  proposal by Oono and Paniconi and using results of the macroscopic
  fluctuation theory, we give a natural definition of a renormalized
  work performed along any given transformation.  We then show that
  the renormalized work satisfies a Clausius inequality and prove that
  equality is achieved for very slow transformations, that is in the
  quasi static limit.  We finally connect the renormalized work to the
  quasi potential of the macroscopic fluctuation theory, that gives
  the probability of fluctuations in the stationary nonequilibrium
  ensemble.
\end{abstract}

%
%



\maketitle

%
%

A main goal of nonequilibrium thermodynamics is to construct analogues
of thermodynamic potentials for nonequilibrium stationary states. 
These potentials should describe the typical macroscopic behavior of
the system as well as the asymptotic probability of fluctuations.  As
it has been shown in \cite{primo}, this program can be implemented
without the explicit knowledge of the stationary ensemble and requires
as input the macroscopic dynamical behavior of systems which can be
characterized by the transport coefficients.  This theory, now known
as \emph{macroscopic fluctuation theory}, is based on an extension of
Einstein equilibrium fluctuation theory to stationary nonequilibrium
states combined with a dynamical point of view. It has been very
powerful in s\-tu\-dy\-ing concrete microscopic models but can be used also
as a phenomenological theory.  It has led to several new interesting
predictions \cite{cumulanti,curr,lagpt,rev,bd,dd}.

From a thermodynamic viewpoint, the analysis of transformations from
one state to another one is most relevant.
This issue has been addressed by several authors in different
contexts. For instance, following the basic papers \cite{J,C,hs}, the
case of Hamiltonian systems with finitely many degrees of freedom has
been recently discussed in \cite{THD,EB} while the case of Langevin
dynamics is considered in \cite{DL}.

We here consider thermodynamic transformations for driven diffusive
systems in the framework of the macroscopic fluctuation theory.  
With respect to the authors mentioned above, the main difference is
that we deal with systems with infinitely many degrees of freedom and
the spatial structure becomes relevant.  For simplicity of notation,
we restrict to the case of a single conservation law, e.g., the
conservation of the mass.  We thus consider an open system in contact
with boundary reservoirs, characterized by their chemical potential
$\lambda$, and under the action of an external field $E$.  We denote
by $\Lambda \subset \bb R^d$ the bounded region occupied by the
system, by $x$ the macroscopic space coordinates and by $t$ the
macroscopic time.  With respect to our previous work
\cite{primo,curr,rev},
we here consider the case in which $\lambda$ and $E$ depend explicitly
on the time $t$, driving the system from a nonequilibrium state to
another one.  The macroscopic dynamics is given by the hydrodynamic
equation for the density which satisfies the following general
assumptions based on the notion of local equilibrium. For stochastic
lattice gases these assumptions can be proven rigorously and the
macroscopic transport coefficient can be characterized in terms of the
underlying microscopic dynamics \cite{S}.

The macroscopic state is completely described by the local density
$u(t,x)$ and the associated current $j(t,x)$. In the sequel we drop
the dependence on the space coordinate $x$ from the notation.  The
macroscopic evolution is given by the continuity equation
\begin{equation}
\label{2.1}
\partial_t u (t) + \nabla\cdot j (t) = 0,
\end{equation}
together with the constitutive equation $j(t)=J(t,u(t))$ expressing the
local current in function of the local density.
For driven diffusive systems the constitutive equation takes the form
\begin{equation}
\label{2.2}
J(t,\rho)  = - D(\rho) \nabla\rho + \chi(\rho) \, E(t)
\end{equation}
where the \emph{diffusion coefficient} $D(\rho)$ and the \emph{mobility}
$\chi(\rho)$ are positive matrices. In the case of time
independent driving the right hand side does not depend explicitly on
time and we denote the current simply by $J(\rho)$.
The transport coefficients in \eqref{2.2} satisfy the 
local Einstein relation $D(\rho) = \chi(\rho) \, f''(\rho)$,
where $f$ is the equilibrium free energy per unit of volume.
The interaction with the external reservoirs specify the boundary
conditions for the evolution defined by \eqref{2.1}--\eqref{2.2}. 
Recalling that $\lambda(t)$ is the chemical potential of the
reservoirs, this boundary condition reads $f'\big(u(t,x) \big) = \lambda(t,x)$,
$x\in\partial \Lambda$.  We shall also assume that when $\lambda$ and
$E$ do not depend on time there is a unique and globally attractive
stationary solution for the flow defined by \eqref{2.1}--\eqref{2.2}
that is denoted by $\bar\rho=\bar\rho_{\lambda,E}$. In particular,
$\bar\rho_{\lambda,E}$ is the typical density profile in the
stationary nonequilibrium state corresponding to time independent
chemical potential $\lambda$ and external field $E$.

Fix time dependent paths $\lambda(t)$ of the chemical potential and
$E(t)$ of the driving field. Given a density profile $\rho$, denote 
by $(u(t),j(t))$, $t \ge 0$,  the solution of \eqref{2.1}--\eqref{2.2}
with initial condition $\rho$.  Let $W_{[0,T]} =
W_{[0,T]}({\lambda, E, \rho})$ be the energy exchanged between the
system and the external driving in the time interval $[0,T]$, that is
\begin{equation}
\label{W=}
W_{[0,T]}=
\int_{0}^{T}\! dt\, \Big\{ 
\int_\Lambda \!dx\: j(t) \cdot E(t) 
- \int_{\partial\Lambda} \!d\sigma \: \lambda (t) \: j(t) \cdot
\hat{n}  
\Big\},
\end{equation}
where $\hat n$ is the outer normal to $\partial\Lambda$ and $d\sigma$
is the surface measure on $\partial \Lambda$. 
The first term on the
right hand side is the energy provided by the external field
while the second is the energy provided by the reservoirs.

In view of the boundary conditions and the Einstein
relation, by using the divergence theorem in
\eqref{W=}, we deduce that 
\begin{equation}
\label{04}
\begin{split}
W_{[0,T]} = & \; F (u(T)) - F(\rho)  
\\
&+ \int_{0}^{T} \!dt  \int_\Lambda\!dx \, j(t)\cdot \chi(u(t) )^{-1} j(t),
\end{split}
\end{equation}
where $F$ is the equilibrium free energy functional,
\begin{equation}
\label{10}
F(\rho) = \int_\Lambda \!dx \: f (\rho(x)).
\end{equation}

Consider two stationary states corresponding to (time-independent)
$(\lambda_0,E_0)$ and $(\lambda_1,E_1)$ and denote by $\bar\rho_0
=\bar\rho_{\lambda_0,E_0}$ and $\bar\rho_1 =\bar\rho_{\lambda_1,E_1}$
the associated density profiles. Such states can be either equilibrium
or nonequilibrium states.
We can drive the system from the initial state $\bar\rho_0$ at time $t=0$ to
the final state $\bar\rho_1$ at time $t=+\infty$ by considering a time
dependent forcing $(\lambda(t),E(t))$ satisfying
$(\lambda(0),E(0))=(\lambda_0,E_0)$ and
$(\lambda(+\infty),E(+\infty))=(\lambda_1,E_1)$. As the second term on
the right hand side of \eqref{04} is positive, by letting
$W=W_{[0,+\infty)}$ be the total energy exchanged in the
transformation, we deduce the Clausius inequality
\begin{equation}
  \label{ci}
  W\ge \Delta F = F(\bar\rho_1)-F(\bar\rho_0).
\end{equation}

When the initial and final states are equilibrium states, e.g., the
external field $E$ vanishes and the chemical potential $\lambda$ is
constant, the inequality \eqref{ci} is a standard formulation
of the second law of thermodynamics. Moreover, by considering  a sequence
of transformations in which the variation of the driving becomes
very slow, it is not difficult to show that equality in
\eqref{ci} is achieved in the quasi static limit, we refer to
\cite{tdmf} for the details. 
On the other hand, for nonequilibrium states
the inequality \eqref{ci} does not carry any information. 
Indeed, as nonequilibrium states support a non vanishing current, in
the limit $T\to+\infty$ the second term on the right hand side of
\eqref{04} becomes infinite so that the left hand side of \eqref{ci}
is infinite  while the right hand side is bounded.  By interpreting the
ideas in \cite{op}, further developed in \cite{hs,knst}, we next
define a renormalized work for which a 
significant Clausius inequality can be obtained also for nonequilibrium
stationary states.

To this aim, we recall the quasi potential, which is the key notion of
the macroscopic fluctuation theory. Consider a system with time
independent driving and let $(\hat u(t), \hat \jmath (t))$, $t\in
[T_1,T_2]$ be a pair density--current satisfying the
continuity equation $\partial_t\hat u +\nabla\cdot \hat\jmath =0$.
According to the basic principles of the macroscopic fluctuation theory
\cite{primo,curr,rev}, the probability of observing this path is given, up to a
prefactor, by
$\exp\big\{ - \epsilon^{-d} \,\beta \, I_{[T_1,T_2]}(\hat u,
\hat\jmath) \big\}$
where $\epsilon$ is the scaling parameter, i.e., the ratio between the
microscopic length scale (say the typical intermolecular distance) and
the macroscopic one, $\beta=1/\kappa T$ (here $T$ is the temperature
and $\kappa$ is Boltzmann's constant), and the action functional
$I$ has the form
\begin{equation}
  \label{af}
  \begin{split}
  I_{[T_1,T_2]}(\hat u, \hat\jmath) 
  &=
  \frac 14 \int_{T_1}^{T_2}\!dt\!\int_{\Lambda}\!dx \,
  \big[\hat\jmath(t) - J(\hat u(t)) \big] 
  \\
  & \qquad\qquad 
  \cdot \chi(\hat u(t))^{-1} \big[\hat\jmath(t) - J(\hat u(t)) \big].  
  \end{split}
\end{equation}
In particular, if $(\hat u,\hat\jmath)$ solves
\eqref{2.1}--\eqref{2.2} then $I_{[T_1,T_2]}(\hat u, \hat\jmath)=0$.
The above statement therefore implies that the typical behavior of
the system is described by the hydrodynamic equations.
The quasi potential is the functional on the set of density
profiles  defined by the variational problem 
\begin{equation}
  \label{qp}
  V(\rho) = \inf \Big\{ I_{(-\infty,0]}(\hat u, \hat\jmath) \,,\:
    \hat u(0) =\rho \Big\}
\end{equation}
where the infimum is carried out over all the trajectories satisfying
the prescribed boundary condition.  Namely, $V(\rho)$ is the minimal
action to bring the system from the typical density profile $\bar\rho$
to the fluctuation $\rho$.  The probability of a density profile
$\rho$ in the stationary nonequilibrium ensemble is then given, up to
a prefactor, by $\exp\big\{ - \epsilon^{-d} \,\beta \, V(\rho) \big\}$.
In particular, the minimizer of $V$ is the typical density profile
$\bar\rho$. 
For equilibrium states it can be shown \cite{rev} that $V$ coincides,
apart an affine transformation, with the free energy functional
\eqref{10}. Moreover, as shown in \cite{primo}, the functional $V$ solves
the stationary Hamilton-Jacobi equation 
\begin{equation}
\label{30}
  \int_{\Lambda} \!dx\, \nabla  
  \frac {\delta V}{\delta\rho} \cdot \chi(\rho) \, \nabla  
  \frac {\delta V}{\delta\rho} 
  - \int_{\Lambda}\!dx\, \frac {\delta V}{\delta\rho}
  \, \nabla \cdot J(\rho) = 0 
\end{equation}
where ${\delta V}/{\delta \rho}$ vanishes at the boundary
$\partial \Lambda$ and $\rho$ satisfies the boundary condition
$f'(\rho(x)) = \lambda(x)$, $x\in\partial\Lambda$.

In the case of time independent driving, the current $J(\rho)$ in
\eqref{2.2} can be decomposed as $J(\rho) = J_{\mathrm{S}}(\rho) +
J_{\mathrm{A}}(\rho)$, where $J_{\mathrm{S}}(\rho)= -\chi(\rho) \,
\nabla \frac {\delta V}{\delta\rho}$ and $J_{\mathrm{A}}(\rho) =
J(\rho) - J_{\mathrm{S}}(\rho)$.  In view of the stationary
Hamilton-Jacobi equation \eqref{30}, the above decomposition is
orthogonal in the sense that for each density profile $\rho$
\begin{equation}
\label{ort}
\int_{\Lambda}\!dx \: J_{\mathrm{S}}(\rho) \cdot \chi(\rho)^{-1}
J_{\mathrm{A}} (\rho) = 0 .
\end{equation}
We shall refer to $J_{\mathrm{S}}(\rho)$ as the \emph{symmetric}
current and to $J_{\mathrm{A}}(\rho)$ as the \emph{antisymmetric}
current. This terminology refers to symmetric and antisymmetric part
of the microscopic dynamics \cite{primo,curr}.  We remark that
$J_{\mathrm{S}}$ is proportional to the thermodynamic force and the
above decomposition depends on the external driving.

Since the quasi potential $V$ is minimal in the stationary profile,
the above definitions imply that $J_{\mathrm{S}} (\bar\rho)=0$;
namely, the stationary current is purely antisymmetric.  In
particular, $J_{\mathrm{A}} (\bar\rho)$ is the typical current in the
stationary nonequilibrium ensemble associated and it is therefore
experimentally accessible.  In view of the general formula \eqref{04}
for the total work, the amount of energy per unit of time needed to
maintain the system in the stationary profile $\bar\rho$ is
\begin{equation}
\label{toman}
\int_\Lambda \!dx \: 
J_\mathrm{A}(\bar\rho) \cdot \chi(\bar\rho)^{-1} 
J_\mathrm{A}(\bar\rho).
\end{equation}

We shall next consider time dependent driving and define a
renormalized work by subtracting from the total work  
the energy needed to maintain the system out of equilibrium.
Fix, therefore,
$T>0$, a density profile $\rho$, and space-time dependent chemical
potentials $\lambda(t)$ and external field $E(t)$, $t\in [0,T]$. 
Let $(u(t),j(t))$ be the corresponding solution of 
\eqref{2.1}--\eqref{2.2} with initial condition $\rho$.
Recalling \eqref{toman}, we define the renormalized work
$W^\textrm{ren}_{[0,T]} = W^\textrm{ren}_{[0,T]}(\lambda,E,\rho)$
performed by the reservoirs and the external field in the time
interval $[0,T]$ as
\begin{equation}
\label{Weff}
\begin{split}
&W^\textrm{ren}_{[0,T]} = W_{[0,T]} 
\\
&\quad 
- \int_{0}^{T}\! dt \int_\Lambda \!dx\, J_\mathrm{A}(t,u(t)) \,
\cdot \chi(u(t))^{-1} J_\mathrm{A}(t,u(t)).
\end{split}
\end{equation}
In this formula $W_{[0,T]} =W_{[0,T]} (\lambda,E,\rho)$ is given in \eqref{W=},
\begin{equation*}
  \begin{split}
  &  J(t,u) =  J_\mathrm{S}(t,u) +J_\mathrm{A}(t,u),
  \\
  & J_\mathrm{S}(t,u) = - \chi(u) \, \nabla  
  \frac {\delta V_{\lambda(t),E(t)}(u)}{\delta u}
  \end{split}
\end{equation*}
in which $u$ is a generic density profile, $J(t,u)$ is given by \eqref{2.2}, and 
$V_{\lambda(t),E(t)}$ is the quasi potential relative to
the state $(\lambda(t), E(t))$ with frozen $t$.
Observe that the definition of the renormalized work involves the
antisymmetric current $J_\mathrm{A}(t)$ computed not at density
profile $\bar\rho_{\lambda(t), E(t)}$ but at the solution
$u(t)$ of the time dependent hydrodynamic equation.
That is, at time $t$ we subtract the power the system would have
dissipated if its actual state $u(t)$ had been the stationary profile
corresponding to $(\lambda(t), E(t))$.  
This choice, which is certainly reasonable for slow transformations,
leads to a Clausius inequality.  
Indeed, by using \eqref{04} and the orthogonality between 
the symmetric and the antisymmetric part of the current,
\begin{equation*}
  \begin{split}
    & W^\textrm{ren}_{[0,T]} (\lambda,E,\rho) = F(u(T)) - F(\rho) 
    \\ &\quad 
    + \int_{0}^{T}\! dt \int_\Lambda \!dx\, J_\mathrm{S}(t,u(t)) \cdot
    \chi(u(t))^{-1} J_\mathrm{S}(t,u(t)).
  \end{split}
\end{equation*}
Consider a density profile $\rho$ and a space-time dependent chemical
potential and external field $(\lambda(t),E(t))$, $t\ge 0$, converging
to $(\lambda_1,E_1)$ as $t\to +\infty$.  
Let $\bar\rho_1 = \bar \rho_{\lambda_1,E_1}$ be the stationary profile 
associated to $(\lambda_1, E_1)$ and $(u(t),j(t))$, $t \ge 0$, be
the solution of \eqref{2.1}--\eqref{2.2} with initial condition
$\rho$.  Since $u(T)$ converges to $\bar\rho_1$, the symmetric part of
the current, $J_\mathrm{S}(u(T))$, relaxes as $T\to +\infty$ to
$J_\mathrm{S}(\bar\rho_1) = 0$.
Under suitable assumptions on the transformation, 
the last integral in the previous formula is convergent
as $T\to+\infty$. By letting 
$W^\textrm{ren}= \lim_{T\to\infty} W^\textrm{ren}_{[0,T]}$, we thus
get 
\begin{equation}
\label{16}
\begin{split}
  & W^\textrm{ren} (\lambda,E,\rho) =  F(\bar\rho_1) - F(\rho) 
  \\
  &\quad + \int_{0}^{\infty} \!dt \int_\Lambda \!dx \:
  J_\mathrm{S}(t,u(t)) \cdot \chi(u(t))^{-1} J_\mathrm{S}(t,u(t))
\end{split}
\end{equation}
where $F$ is the equilibrium free energy functional \eqref{10}.
In particular,
\begin{equation}
\label{15}
W^\textrm{ren}(\lambda, E,\rho)  \ge  F (\bar\rho_1) - F(\rho),
\end{equation}
which is a meaningful version of the Clausius inequality for
nonequilibrium states.  
Furthermore, by considering a sequence of transformations $(\lambda(t), E(t))$
which vary on a time scale $1/\delta$, we realize that the integrand
on second term in the right hand side of \eqref{16} is of order
$\delta^2$ while the integral essentially extends, due to the finite 
relaxation time of the system, over an interval of order $\delta^{-1}$.
Therefore in quasi static limit $\delta \to 0$ equality in \eqref{15}
is achieved. We refer to \cite{tdmf} for more details.

For special transformations, the integral in \eqref{16}, which
represents the \emph{excess work}  over a quasi static
transformations, can be related to the quasi potential. 
Consider at time $t=0$ a stationary nonequilibrium profile
$\bar\rho_0$ corresponding to some driving $(\lambda_0,E_0)$.  
The system is put in contact with new reservoirs at chemical potential
$\lambda_1$ and a new external field $E_1$.  For $t> 0$ the system
evolves according to the hydrodynamic equation
\eqref{2.1}--\eqref{2.2} with initial condition $\bar\rho_0$, time
independent boundary condition $\lambda_1$ and external field $E_1$.
In particular, as $t\to +\infty$ the system relaxes to $\bar\rho_1$,
the stationary density profile corresponding to $(\lambda_1,E_1)$.
A simple calculation shows that along such a path
\begin{equation}
\label{u=qp}
\begin{split}
\int_{0}^{\infty} \!dt \int_\Lambda \!dx \:
  J_\mathrm{S}(t,u(t)) \cdot \chi(u(t))^{-1} J_\mathrm{S}(t,u(t))\\
=V_{\lambda_1,E_1} (\bar\rho_0)
- V_{\lambda_1,E_1} (\bar\rho_1) 
= V_{\lambda_1,E_1} (\bar\rho_0).
\end{split}
\end{equation}
The quasi potential  $V_{\lambda_1,E_1} (\bar\rho_0)$ 
thus represents the excess
work, with respect to a quasi static transformation, 
along the path that solves \eqref{2.1}--\eqref{2.2} with initial
condition $\bar\rho_0$ and time-independent driving $(\lambda_1, E_1)$.

To connect the above result with classical equilibrium thermodynamics,
consider an equilibrium state with vanishing external field and
constant chemical potential $\lambda_0$ and let $\bar\rho_0$ be the
corresponding homogeneous density, i.e.\ $\lambda_0 = f'(\bar\rho_0)$.
The system is put in contact with a new environment with chemical
potential $\lambda_1$. In this case, recalling that $f$ is the free
energy per unit of volume and that the temperature of the system is
the same of the environment, the \emph{avalaibility} per unit of
volume is defined, see \cite[Ch.~7]{pippard}, by $a= f(\bar\rho_0)
-\lambda_1 \bar\rho_0$.  The function $a$, which depends on state of
the system and the environment, can be used to compute the maximal
useful work that can be extracted from the system in the given
environment. More precisely, by letting $\bar\rho_1$ be such that
$f'(\bar\rho_1)=\lambda_1$, then
$-\Delta a = f(\bar\rho_0) - f(\bar\rho_1) - \lambda_1 (\bar\rho_0
-\bar\rho_1)\ge 0$
is the maximal useful work per unit of volume
that can be extracted from the system in the given environment.
By computing the quasi potential for equilibrium
states, see \cite{rev}, 
we get $V_{\lambda_1,0}(\bar\rho_0) = - |\Lambda| \Delta a$.  
Therefore, while a
definition of thermodynamic potentials, that is functionals of the
state of the system, does not appear possible in nonequilibrium
thermodynamics, the quasi potential is the natural extension of the
availability.

In terms of the underlying microscopic ensembles, as discussed in
\cite{tdmf}, the quasi potential $V_{\lambda_1,E_1} (\bar\rho_0)$ can
be obtained by computing the relative entropy of the ensemble
associated to $(\lambda_0, E_0)$ with respect to the one  associated 
to $(\lambda_1, E_1)$.
By considering a Markovian model for such underlying dynamics, it is
also possible to give a microscopic definition of the exchanged work
which in the hydrodynamic scaling limit converges to \eqref{W=}. The
corresponding fluctuations can be deduced from those of the
empirical current \cite{curr}.

The definition of renormalized work we have introduced is natural and
ensures, as we have discussed, both its finiteness and the validity of
a Clausius inequality.  From an operational point of view, the quasi
potential, a generically nonlocal quantity, can be obtained from the
measurement of the density correlation functions. In fact $V$ is the
Legendre transform of the generating functional of density correlation
functions \cite{primo}.  On the other hand, the identity
$W^\textrm{ren} = \Delta F$, that is achieved for quasi static
transformations, requires the knowledge of the total current in the
intermediate stationary states that can be directly measured.
 
One may ask whether there exist, with respect to \eqref{Weff},
alternative renormalizations of the total work. For instance, 
in the recent work \cite{copioni}, Maes and Netocny considered the
topic of a renormalized Clausius inequality in the
context of a single Brownian particle in a time dependent environment.
To compare the approach in \cite{copioni} to the present one, consider
$N$ independent diffusions in the thermodynamic limit $N\to \infty$. 
Each diffusion solves the Langevin equation $\dot X =
E(t,X) +\sqrt{2} \, \dot w$, where $E$ is a time dependent vector
field and $\dot w$ denotes white noise. The corresponding 
stationary measure with $E$ frozen at time $t$ is denoted 
by $\exp\{-v(t,x)\}$. 
The scheme discussed here can be now applied, the hydrodynamic
equations are \eqref{2.1}--\eqref{2.2} with $D=1$ and
$\chi(\rho)=\rho$. Our renormalized work is given by \eqref{Weff} with 
$J_\mathrm{A}(t,\rho)= \rho \big[ E(t,x) + \nabla v(t,x) \big]$.
The renormalization introduced in \cite{copioni} is instead obtained
by introducing a potential field such that the corresponding
stationary state has minimal entropy production. Namely, they write 
$E= f -\nabla U$ and subtract from the energy exchanged the space-time
integral of $|J_t^\phi|^2/\rho$ where $J_t^\phi= \rho (f -\nabla \phi)-\nabla
\rho$ and $\phi=\phi(t,x;\rho)$ is chosen so that $\nabla\cdot J_t^\phi=0$.
While the two renormalization schemes are different, 
both satisfy the Clausius inequality \eqref{15} with $F(\rho)=\int\!dx\,
\rho\log\rho$.  Observe that in this case of independent particles our
renormalization is local while the dependence of $J_t^\phi$ on $\rho$ is
nonlocal.
It is not clear to us how the approach in \cite{copioni} can be
generalized to cover the case of interacting particles in the
hydrodynamic scaling limit.

We are in debt to J.L.\ Lebowitz for continued discussions on the topics
here considered. We also acknowledge stimulating exchanges of views
with C.\ Di Castro, T.\ Komatsu, N.\ Nakagawa, S.\ Sasa, and H.\
Tasaki.

\end{document}